\documentclass[sigconf]{acmart}

\def\BibTeX{{\rm B\kern-.05em{\sc i\kern-.025em b}\kern-.08emT\kern-.1667em\lower.7ex\hbox{E}\kern-.125emX}}

\copyrightyear{2019}
\acmYear{2019}
\setcopyright{acmlicensed}
\acmConference[KDD '19]{KDD '19}{August 4-8, 2019}{Anchorage Alaska, USA}
\acmPrice{15.00}
\acmDOI{10.1145/1122445.1122456}

\usepackage{amsmath}
\usepackage{graphicx}
\usepackage{color}
\usepackage[export]{adjustbox}
\usepackage[caption=false]{subfig}
\usepackage{tabularx}
\usepackage{booktabs}
\usepackage{bm}
\usepackage{multirow}
\usepackage{array}
\usepackage{enumitem}
\usepackage{url}
\usepackage{hyperref}
\usepackage{algorithm}
\usepackage{algorithmicx}
\usepackage{algpseudocode}
\algnewcommand\algorithmicinput{\textbf{INPUT:}}
\algnewcommand\INPUT{\item[\algorithmicinput]}
\algnewcommand\algorithmicoutput{\textbf{OUTPUT:}}
\algnewcommand\OUTPUT{\item[\algorithmicoutput]}

\newcommand{\tabincell}[2]{\begin{tabular}{@{}#1@{}}#2\end{tabular}}

\graphicspath{ {.} }

\begin{document}

%
% The "title" command has an optional parameter, allowing the author to define a "short title" to be used in page headers.
\title{Multi-Interest Network with Dynamic Routing for Recommendation at Tmall}

%
% The "author" command and its associated commands are used to define the authors and their affiliations.
% Of note is the shared affiliation of the first two authors, and the "authornote" and "authornotemark" commands
% used to denote shared contribution to the research.

\author{Chao Li}
\authornote{Both authors contributed equally to this research.}
\author{Zhiyuan Liu}
\authornotemark[1]
\email{{ruide.lc,michong.lzy}@alibaba-inc.com}
\affiliation{
  \institution{Alibaba Group}
  \city{Beijing}
  \state{China}
}

\author{Mengmeng Wu}
\author{Yuchi Xu}
\author{Pipei Huang}
\authornote{Pipei Huang is the Corresponding author.}
\email{{max.wmm,yuchi.xyc}@alibaba-inc.com}
\email{pipei.hpp@alibaba-inc.com}
\affiliation{
  \institution{Alibaba Group}
  \city{Beijing}
  \state{China}
}

\author{Huan Zhao}
\email{hzhaoaf@cse.ust.hk}
\affiliation{
  \institution{Department of Computer Science and Engineering}
  \institution{Hong Kong University of Science and Technology}
  \city{Kowloon}
  \country{Hong Kong}
}

\author{Guoliang Kang}
\email{Guoliang.Kang@student.uts.edu.au}
\affiliation{
  \institution{Centre for Artificial Intelligence}
  \institution{University of Technology Sydney}
  \city{Sydney}
  \country{Australia}
}

\author{Qiwei Chen}
\author{Wei Li}
\email{{chenqiwei.cqw,rob.lw}@alibaba-inc.com}
\affiliation{
  \institution{Alibaba Group}
  \city{Hangzhou}
  \country{China}
}

\author{Dik Lun Lee}
\email{dlee@cse.ust.hk}
\affiliation{%
  \institution{Department of Computer Science and Engineering}
  \institution{Hong Kong University of Science and Technology}
  \city{Kowloon}
  \country{Hong Kong}
}

%
% The abstract is a short summary of the work to be presented in the article.
\begin{abstract}

Industrial recommender systems usually consist of the matching stage and the ranking stage,  in order to handle the billion-scale of users and items.
The matching stage retrieves candidate items relevant to user interests, while the ranking stage sorts candidate items by user interests.
Thus, the most critical ability is to model and represent user interests for either stage.
Most of the existing deep learning-based models represent one user as a single vector which is insufficient to capture the varying nature of user's interests.
%Though a few progress has been made for capturing diverse interests of users in the ranking stage, it remains challenging for the matching stage.
In this paper, we approach this problem from a different view, to represent one user with multiple vectors encoding the different aspects of the user's interests.
We propose the Multi-Interest Network with Dynamic routing (MIND) for dealing with user's diverse interests in the matching stage.
Specifically, we design a multi-interest extractor layer based on capsule routing mechanism, which is applicable for clustering historical behaviors and extracting diverse interests.
Furthermore, we develop a technique named label-aware attention to help learn a user representation with multiple vectors.
Through extensive experiments on several public benchmarks and one large-scale industrial dataset from Tmall, we demonstrate that MIND can achieve superior performance than state-of-the-art methods for recommendation.
Currently, MIND has been deployed for handling major online traffic at the homepage on Mobile Tmall App.

\end{abstract}

\settopmatter{printacmref=false} % Removes citation information below abstract
\renewcommand\footnotetextcopyrightpermission[1]{} % removes footnote with conference information in first column
\pagestyle{plain} % removes running headers
\setcopyright{none}

%
% This command processes the author and affiliation and title information and builds
% the first part of the formatted document.
\maketitle

\section{Introduction}

\begin{figure}[ht]
\centering
\includegraphics[scale=0.25]{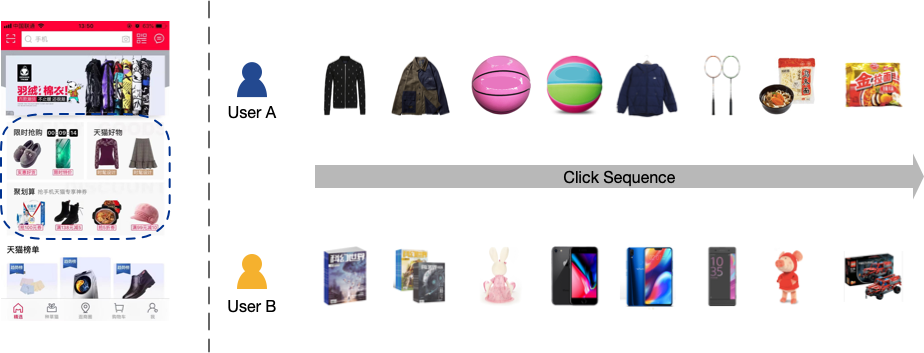}
\caption{Left: The areas highlighted with dashed rectangle are personalized for billion-scale users at Tmall; Right: User A interacts with products from several different categories, including clothes, sports and food, while user B interacts with products of books, toys and cellphones.}
\label{fig:Introduction}
\end{figure}

Tmall, the biggest Business-To-Customer (B2C) e-commerce platform in China, serves billion-scale users by providing billion-scale products online.
On November 11-th of 2018, the well-known Tmall global shopping festival, the Gross Merchandise Volume (GMV) is around 213 billion yuan, achieving an increase rate of 26.9\% compared with the same day of 2017.
As the number of users and products is continuously growing, it becomes increasingly important to help each user find products that he/she might be interested in.
In recent years, Tmall has spent huge efforts in developing personalized recommender systems (RS for short), which significantly contribute to the optimization of user experience and the increase of business value.
For example, the homepage on Mobile Tmall App (as shown in Figure \ref{fig:Introduction} (Left)), which accounts for about half of total traffic at Tmall, has deployed RS for displaying personalized products to meet customers' personalized need.

Due to the billion-scale users and items, the recommendation process designed for Tmall consists of two stages, the matching stage and the ranking stage. The matching stage is responsible for retrieving thousands of candidate items that are relevant to user interests, after which the ranking stage predicts precise probabilities of users interacting with these candidate items.
For both of the two stages, it is vital to model user interests and find user representations capturing user interests, in order to support efficient retrieval of items that satisfy users' interests.
However, it is non-trivial to model user interests at Tmall, due to the existence of diverse interests of users.
On average, billion-scale users visit Tmall, each user interacts with hundreds of products every day.
The interacted products tend to belong to different categories, indicating the diversity of user interests.
For example, as shown in Figure \ref{fig:Introduction} (Right),
different users are distinct in terms of their interests and the same user may also be interested in various kinds of items.
Therefore, the capability of capturing user's diverse interests becomes vital for RS at Tmall.

Existing recommendation algorithms model and represent user interests in different ways.
Collaborative filtering-based methods represent user interests by historical interacted items \cite{sarwar2001item} or hidden factors \cite{koren2009matrix}, which suffer from sparsity problem or computationally demanding.
Deep learning-based methods usually represent user interests with low-dimensional embedding vectors.
For example, the deep neural network proposed for YouTube video recommendation (YouTube DNN) \cite{covington2016deep} represents each user by one fixed-length vector transformed from the
past behaviors of users, which can be a bottleneck for modeling diverse interests, as its dimensionality must be large in order to express the huge number of interest profiles at Tmall.
Deep Interest Network (DIN)\cite{zhou2018deep} makes the user representation vary over different items with attention mechanisms to capture the diversity of user interests.
Nevertheless, the adoption of attention mechanisms also makes it computationally prohibitive for large-scale applications with billion-scale items as it requires re-calculation of user representation for each item, making DIN only applicable for the ranking stage.

In this paper, we focus on the problem of modeling diverse interests of users in the matching stage.
In order to overcome the limitations of existing methods, we propose the Multi-Interest Network with Dynamic routing (MIND) for learning user representations that reflect diverse interests of users in the matching stage of industrial RS.
To infer the user representation vectors, we design a novel layer called multi-interest extractor layer, and this layer utilizes dynamic routing \cite{sabour2017dynamic} to adaptively aggregate user's historical behaviors into user representations.
The process of dynamic routing can be viewed as soft-clustering, which groups user’s historical behaviors into several clusters.
Each cluster of historical behaviors is further used to infer the user representation vector corresponding to one particular interest.
In this way, for a particular user, MIND outputs multiple representation vectors, which collectively represent diverse interests of users.
The user representation vectors are computed only once and can be used in the matching stage for retrieving relevant items from billion-scale items.
To summarize, the main contributions of this work are as follows:
\begin{itemize}
  \item To capture diverse interests of users from user behaviors, we design the multi-interest extractor layer, which utilizes dynamic routing to adaptively aggregate user's historical behaviors into user representation vectors.
  \item By using user representation vectors produced by the multi-interest extractor layer and a newly proposed label-aware attention layer, we build a deep neural network for personalized recommendation tasks. Compared with existing methods, MIND shows superior performance on several public datasets and one industrial dataset from Tmall.
  \item To deploy MIND for serving billion-scale users at Tmall, we construct a system to implement the whole pipeline for data collecting, model training and online serving. The deployed system significantly improves the click-through rate (CTR) of the homepage on Mobile Tmall App.
\end{itemize}
The remainder of this paper is organized as follows:
related works are reviewed in section 2;
Section 3 elaborates the technical details of MIND;
In section 4, we detail the experiments for comparing MIND with existing methods on several public benchmarks and online serving;
Section 5 introduces the deployment of MIND in large-scale industrial application;
The last section gives conclusion and future work of this paper.

\begin{figure*}[ht]
\centering
\includegraphics[scale=0.35]{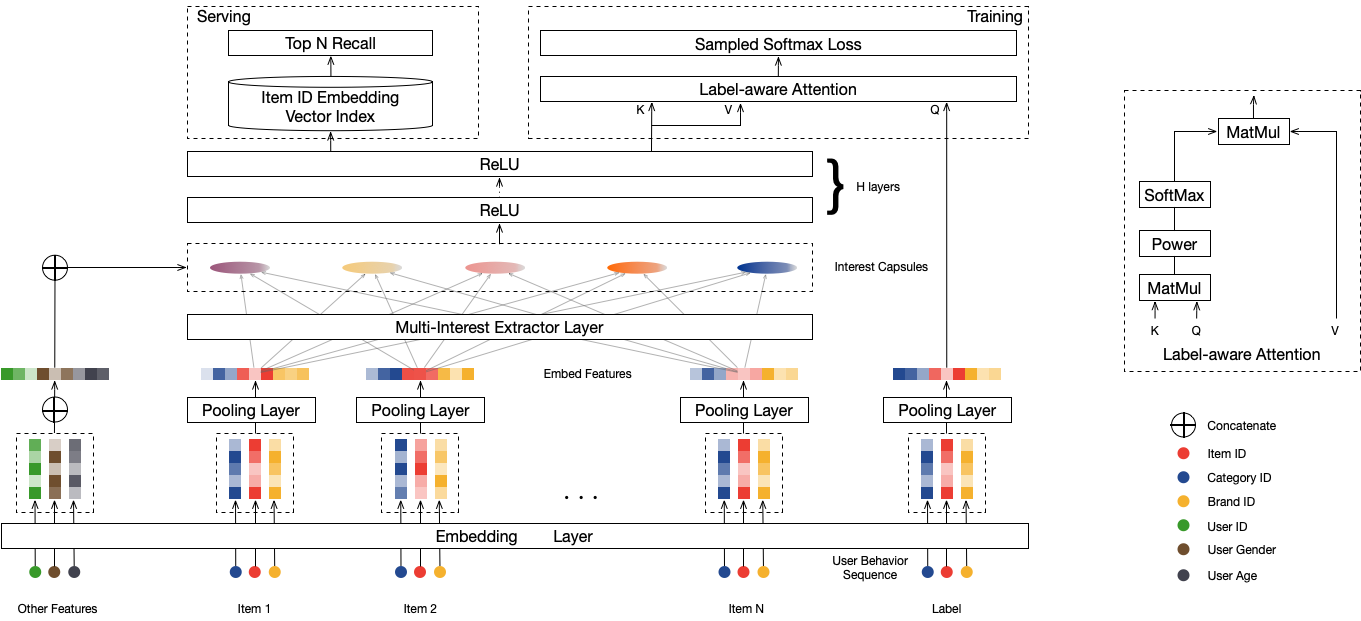}
\caption{Overview of MIND.
MIND takes user behaviors with user profile features as inputs, and outputs user representation vectors for item retrieval in the matching stage of recommendation.
Id features from input layer are transformed into embeddings through the embedding layer, and embeddings of each item are further averaged by a pooling layer.
User behavior embeddings are fed into the multi-interest extractor layer, which produces interest capsules.
By concatenating interest capsules with user profile embedding and transforming the concatenated capsules by several ReLU layers, user representation vectors are obtained.
During training, an extra label-aware attention layer is introduced to guide the training process.
At serving, the multiple user representation vectors are used to retrieve items through an approximate nearest neighbor lookup approach.
}

\label{fig:ModelArchitecture}
\end{figure*}

\section{Related Work}

\textbf{Deep Learning for Recommendation.}
Inspired by the success of deep learning in computer vision and natural language processing \cite{lecun2015deep}, much efforts have been put for developing deep learning-based recommendation algorithms \cite{batmaz2018review}.
Besides the industrial applications proposed by \cite{covington2016deep,zhou2018deep}, various types of deep models have gained significant attention.
Neural Collaborative Filtering (NCF) \cite{he2017neural}, DeepFM \cite{guo2017deepfm} and Deep Matrix Factorization Models (DMF) \cite{xue2017deep} construct a neural network composed of several MLPs to model the interaction between users and items.
\cite{tang2018personalized} presents a novel solution to top-N sequential recommendation by providing an united and flexible network for capturing more features.

\textbf{User Representation.}
Representing users as vectors is commonly used in RS.
Traditional methods assembles user preference as vectors composed of interested items \cite{herlocker2002empirical,bell2007improved,sarwar2001item}, keywords \cite{cantador2010content,elkahky2015multi} and topics \cite{yin2015dynamic}.
As the emergence of distributed representation learning, user embeddings obtained by neural networks are widely used.
\cite{chen2016learning} employs RNN-GRU to learn user embeddings from the temporal ordered review documents.
\cite{yu2016user} learns user embedding vectors from word embedding vectors and applies them to recommending scholarly microblogs.
\cite{amir2016modelling} proposes a novel convolutional neural network based model that explicitly learns and exploits user embeddings in conjunction with features derived from utterances.

\textbf{Capsule Network.}
The concept of "Capsule", a small group of neurons assembled to output a whole vector, is firstly proposed by Hinton \cite{hinton2011transforming} at 2011.
Instead of backpropagation, dynamic routing \cite{sabour2017dynamic} is used to learn the weights on the connections between capsules, which is improved by utilizing Expectation-Maximization algorithm \cite{hinton2018matrix} to overcome several deficiencies and achieves better accuracy.
These two main differences to conventional neural network make capsule networks capable of encoding the relationship between the part and the whole, which is adavanced in computer vision and natural language processing.
SegCaps \cite{lalonde2018capsules} proves that capsules can successfully model the spatial relationships of the objects better than traditional CNNs.
\cite{zhao2018investigating} investigates the capsule networks for text classification and proposes 3 strategies to boost the performance.

\section{Method}

\subsection{Problem Formalization}
The objective of the matching stage for industrial RS is to retrieve a subset of items from the billion-scale item pool $\mathcal{I}$ for each user $u \in \mathcal{U}$ such that the subset contains only thousands of items and each item is relevant to interests of the user.
In order to achieve this objective, historical data generated by RS is collected for building a matching model.
Specifically, each instance can be represented by a tuple $(\mathcal{I}_u, \mathcal{P}_u, \mathcal{F}_i)$, where $\mathcal{I}_u$ denotes the set of items interacted by user $u$ (also called user behavior), $\mathcal{P}_u$ the basic profiles of user $u$ (like user gender and age), $\mathcal{F}_i$ the features of target item (such as item id and category id).

The core task of MIND is to learn a function for mapping raw features into user representations, which can be formulated as
\begin{equation}
  \boldsymbol{\textrm{V}}_u = \emph{f}_{user} \left( \mathcal{I}_u, \mathcal{P}_u \right),
\end{equation}
where $\boldsymbol{\textrm{V}}_u = \left( \overrightarrow{\bm{v}}_u^{1}, ..., \overrightarrow{\bm{v}}_u^{K} \right) \in \mathbb{R}^{d \times K} $ denotes the representation vectors of user $u$, $d$ the dimensionality, $K$ the number of representation vectors.
When $K=1$, one representation vector is used, just like YouTube DNN.
Besides, the representation vector of target item $i$ is obtained by an embedding function as
\begin{equation}
  \overrightarrow{\boldsymbol{e}}_i = \emph{f}_{item} \left( \mathcal{F}_i \right),
\end{equation}
where $\overrightarrow{\boldsymbol{e}}_i \in \mathbb{R}^{d \times 1}$ denotes the representation vector of item $i$, and the detail of $\emph{f}_{item}$ will be illustrated in the "Embedding \& Pooling Layer" section.

When user representation vectors and item representation vector are learned, top $N$ candidate items are retrieved according to the scoring function
\begin{equation}
  \emph{f}_{score} \left(\boldsymbol{\textrm{V}}_u,  \overrightarrow{\boldsymbol{e}}_i \right) = \max \limits_{1\le k \le K}  \overrightarrow{\boldsymbol{e}}_i^{\textrm{T}} \overrightarrow{\bm{v}}_u^{k},
\end{equation}
where $N$ is the predefined number of items to be retrieved in the matching stage.

\subsection{Embedding \& Pooling Layer}
As shown in Figure 2, the input of MIND consists of three groups, user profile $\mathcal{P}_u$, user behavior $\mathcal{I}_u$, and label item $\mathcal{F}_i$.
Each group contains several categorical id features, and these id features are of extremely high dimensionality.
For instance, the number of item ids is about billions, thus we adopt the widely-used embedding technique to embed these id features into low-dimensional dense vectors (a.k.a embeddings), which significantly reduces the number of parameters and eases the learning process.
For id features (gender, age, etc.) from $\mathcal{P}_u$, corresponding embeddings are concatenated to form the user profile embedding $\overrightarrow{\boldsymbol{p}}_u$.
For item ids along with other categorical ids (brand id, shop id, etc.) that have been proved to be useful for cold-start items \cite{wang2018billion} from $\mathcal{F}_i$, corresponding embeddings are further passed through an average pooling layer to form the label item embedding $\overrightarrow{\boldsymbol{e}}_i$.
Lastly, for items from user behavior $\mathcal{I}_u$, corresponding item embeddings are collected to form the user behavior embedding $\boldsymbol{\textrm{E}}_u = \{ \overrightarrow{\boldsymbol{e}}_j \text{ , } j \in \mathcal{I}_u \}$.

\subsection{Multi-Interest Extractor Layer}
We argue that representing user interests by one representation vector can be a bottleneck for capturing diverse interests of users, because we have to compress all information related with diverse interests of users into one representation vector.
Thus, all information about diverse interests of users is mixed together, causing inaccurate item retrieval for the matching stage.
Instead, we adopt multiple representation vectors to express distinct interests of users separately.
By this way, diverse interests of users are considered separately in the matching stage, enabling more accurate item retrieval for every aspect of interests.

To learn multiple representation vectors, we utilize clustering process to group user's historical behaviors into several clusters.
Items from one cluster are expected to be closely related and collectively represent one particular aspect of user interests.
Here, we design the multi-interest extractor layer for clustering historical behaviors and inferring representation vectors for resulted clusters.
Since the design of multi-interest extractor layer is inspired by the recently proposed dynamic routing for representation learning in capsule network \cite{hinton2011transforming,sabour2017dynamic,hinton2018matrix}, we firstly revisit essential basics in order to make this paper self-contained.

\subsubsection{Dynamic Routing Revisit}
We briefly introduce dynamic routing \cite{sabour2017dynamic} for representation learning of capsules, a new form of neural units represented by vectors.
Suppose we have two layers of capsules, and we refer capsules from the first layer and the second layer as low-level capsules and high-level capsules respectively.
The goal of dynamic routing is to compute the values of high-level capsules given the values of low-level capsules in an iterative way.
In each iteration, given low-level capsules $i \in \{ 1,...,m \}$ with corresponding vectors $\overrightarrow{\boldsymbol{c}}_i^{l} \in \mathbb{R}^{N_l \times 1}, i \in \{ 1,...,m \}$ and high-level capsules $j \in \{ 1,...,n \}$ with corresponding vectors $\overrightarrow{\boldsymbol{c}}_j^{h} \in \mathbb{R}^{N_h \times 1}, j \in \{ 1,...,n \} $, the routing logit $b_{ij}$ between low-level capsule $i$ and high-level capsule $j$ is computed by
\begin{equation}
b_{ij} = (\overrightarrow{\boldsymbol{c}}_j^{h})^T \boldsymbol{\textrm{S}}_{ij} \overrightarrow{\boldsymbol{c}}_i^{l},
\end{equation}
where $\boldsymbol{\textrm{S}}_{ij} \in \mathbb{R}^{N_h \times N_l}$ denotes the bilinear mapping matrix to be learned.

With routing logits calculated, the candidate vector for high-level capsule $j$ is computed as weighted sum of all low-level capsules
\begin{equation}
\overrightarrow{\boldsymbol{z}}_j^{h} = \sum_{i=1}^{m} w_{ij} \boldsymbol{\textrm{S}}_{ij} \overrightarrow{\boldsymbol{c}}_i^{l},
\end{equation}
where $w_{ij}$ denotes the weight for connecting low-level capsule $i$ and high-level capsule $j$ and is calculated by performing softmax on routing logits as
\begin{equation}
w_{ij} = \frac{\exp{b_{ij}}}{\sum_{k=1}^{m} \exp{b_{ik}}}.
\end{equation}

Finally, a non-linear "squash" function is applied to obtain the vectors of high-level capsules as
\begin{equation}
\overrightarrow{\boldsymbol{c}}_{j}^{h} = squash(\overrightarrow{\boldsymbol{z}}_j^{h}) = \frac{\left\lVert \overrightarrow{\boldsymbol{z}}_j^{h} \right\rVert ^ 2}{1 + \left\lVert \overrightarrow{\boldsymbol{z}}_j^{h} \right\rVert ^ 2} \frac{\overrightarrow{\boldsymbol{z}}_j^{h}}{\left\lVert \overrightarrow{\boldsymbol{z}}_j^{h} \right\rVert}.
\end{equation}

The values of $b_{ij}$ are initialized to zeros, and the routing process is usually repeated three times to converge.
When routing finished, high-level capsule's values $\overrightarrow{\boldsymbol{c}}_j^h$ are fixed and can be used as inputs for next layers.\

\subsubsection{B2I Dynamic Routing}
\label{sec:b2i}
In a nutshell, capsule is a new kind of neuron represented by one vector instead of one scalar used in ordinary neural networks.
The vector-based capsule is expected to be able to represent different properties of an entity, in which the orientation of a capsule represents one property and the length of the capsule is used to represent the probability that the property exists.
Correspondingly, the objective of the multi-interest extractor layer is to learn representations for expressing properties of user interests as well as whether corresponding interests exist.
The semantic connection between capsules and interest representations motivates us to regard the behavior/interest representations as behavior/interest capsules and employ dynamic routing to learn interest capsules from behavior capsules.
Nevertheless, the original routing algorithm proposed for image data is not directly applicable for processing user behavior data.
So, we propose Behavior-to-Interest (B2I) dynamic routing for adaptively aggregating user's behaviors into interest representation vectors, and it differs from original routing algorithm in three aspects.

\textit{Shared bilinear mapping matrix.}
We use fixed bilinear mapping matrix $\boldsymbol{\textrm{S}}$ instead of a separate bilinear mapping matrix for each pair of low-level capsules and high-level capsules in original dynamic routing due to two considerations.
On the one hand, user behaviors are of variable-length, ranging from dozens to hundreds for Tmall users, thus the use of fixed bilinear mapping matrix is generalizable.
On the other hand, we hope interest capsules lie in the same vector space, but different bilinear mapping matrice would map interest capsules into different vector spaces.
Thus, the routing logit is calculated by
\begin{equation}
b_{ij} = \overrightarrow{\boldsymbol{u}}_j^T \boldsymbol{\textrm{S}} \overrightarrow{\boldsymbol{e}}_i, \qquad i \in \mathcal{I}_{u}, j \in \{1,...,K \},
\end{equation}
where $\overrightarrow{\boldsymbol{e}}_i \in \mathbb{R}^{d}$ denotes the embedding of behavior item $i$, $\overrightarrow{\boldsymbol{u}}_j \in \mathbb{R}^{d}$ the vector of interest capsule $j$. The bilinear mapping matrix $\boldsymbol{\textrm{S}} \in \mathbb{R}^{d \times d}$ is shared across each pair of behavior capsules and interest capsules.

\textit{Randomly initialized routing logits.}
Owing to the use of shared bilinear mapping matrix $\boldsymbol{\textrm{S}}$, initializing routing logits to zeros will lead to the same initial interest capsules.
Then, the subsequent iterations will be trapped in a situation, where different interest capsules remain the same all the time.
To mitigate this phenomenon, we sample a random matrix from gaussian distribution $\mathcal N(0, \sigma ^ 2)$ for initial routing logits to make initial interest capsules differ from each other, similar to the well-established K-Means clustering algorithm.

\textit{Dynamic interest number.}
As the number of interest capsules owned by different users may be different, we introduce a heuristic rule for adaptively adjusting the value of $K$ for different users.
Specifically, the value of $K$ for user $u$ is computed by
\begin{equation}
\label{adaptive}
K_{u}^\prime = \max(1, \min(K, \log_2 (|\mathcal{I}_{u}|))).
\end{equation}

This strategy for adjusting the number of interest capsules can save some resources, including both computing and memory resources, for those users with fewer interests.

The whole dynamic routing procedure is listed in Algorithm \ref{alg:routing}.

\begin{algorithm}[htb]
\caption{B2I Dynamic Routing.}
\label{alg:routing}
\begin{flushleft}
\textbf{Input:} behavior embeddings $\left\{ \overrightarrow{\boldsymbol{e}}_i, i \in \mathcal{I}_u \right\}$, iteration times $r$, number of interest capsules $K$ \\
\textbf{Output:} interest capsules $\left\{\overrightarrow{\boldsymbol{u}}_j, j=1,...,K_{u}^\prime\right\}$
\end{flushleft}
\begin{algorithmic}[1]
\State calculate adaptive number of interest capsules $K_{u}^\prime$ by \eqref{adaptive}
\State for all behavior capsule $i$ and interest capsule $j$: initialize $b_{ij} \sim \mathcal N(0, \sigma ^ 2)$.
\For {$k \leftarrow 1, r$}
\State for all behavior capsule $i$: $w_{ij} \leftarrow softmax(b_{ij})$
\State for all interest capsule $j$: $\overrightarrow{\boldsymbol{z}}_j = \sum_{i\in \mathcal{I}_u} w_{ij} \boldsymbol{\textrm{S}} \overrightarrow{\boldsymbol{e}}_i$
\State for all interest capsule $j$: $\overrightarrow{\boldsymbol{u}}_j \leftarrow squash(\overrightarrow{\boldsymbol{z}}_j)$
\State for all behavior capsule $i$ and interest capsule $j$: $b_{ij} \leftarrow b_{ij} + \overrightarrow{\boldsymbol{u}}_j^T \boldsymbol{\textrm{S}} \overrightarrow{\boldsymbol{e}}_i$
\EndFor
\State \Return{$\left\{\overrightarrow{\boldsymbol{u}}_j, j=1,...,K_{u}^\prime\right\}$}
\end{algorithmic}
\end{algorithm}

\subsection{Label-aware Attention Layer}
Through multi-interest extractor layer, several interest capsules are generated from user's behavior embeddings.
Different interest capsules represent different aspects of user interests, and the relevant interest capsule is used for evaluating user's preference on specific items.
Therefore, during training, we design a label-aware attention layer based on scaled dot-product attention \cite{vaswani2017attention} to make the target item choose which interest capsule is used.
Specifically, for one target item, we calculate the compatibilities between each interest capsule and target item embedding, and compute a weighted sum of interest capsules as user representation vector for the target item, where the weight for one interest capsule is determined by corresponding compatibility.
In label-aware attention, the label is the query and the interest capsules are both keys and values, as shown in Figure \ref{fig:ModelArchitecture}.
The output vector of user $u$ with respect to item $i$ is computed as
\begin{align*}
\overrightarrow{\boldsymbol{v}}_u
&= \text{Attention} \left( \overrightarrow{\boldsymbol{e}}_i, \boldsymbol{\textrm{V}}_u, \boldsymbol{\textrm{V}}_u \right) \\
&=  \boldsymbol{\textrm{V}}_u \text{softmax}(\text{pow}(\boldsymbol{\textrm{V}}_u^{\textrm{T}}\overrightarrow{\boldsymbol{e}}_i, p)),
\end{align*}
where $\text{pow}$ denotes element-wise exponentiation operator, $p$ a tunable parameter for adjusting the attention distribution.
When $p$ is close to 0, each interest capsule attends to receive even attention.
When $p$ is bigger than 1, as $p$ increases, the value has bigger dot-product will receive more and more weight.
Consider the limit case, when $p$ gets infinity, the attention mechanism becomes a kind of hard attention to pick the value who has the biggest attention and ignore others.
In our experiments, we find out that using hard attention leads to faster convergence.

\subsection{Training \& Serving}
With the user vector $\overrightarrow{\boldsymbol{v}}_u$ and the label item embedding $\overrightarrow{\boldsymbol{e}}_i$ ready, we compute the probability of the user $u$ interacting with the label item $i$ as
\begin{equation}
  \label{next_item_prob}
  \text{Pr} (i|u) = \text{Pr} \left(\overrightarrow{\boldsymbol{e}}_i | \overrightarrow{\boldsymbol{v}}_u \right)
  = \frac{\exp\left(
  \overrightarrow{\boldsymbol{v}}_u^{\textrm{T}} \overrightarrow{\boldsymbol{e}}_i
  \right)
  }
  {
  \sum_{j\in \mathcal{I}} \exp\left(
  \overrightarrow{\boldsymbol{v}}_u^{\textrm{T}} \overrightarrow{\boldsymbol{e}}_j
  \right)
  }.
\end{equation}
Then, the overall objective function for training MIND is
\begin{equation}
  L = \sum_{(u,i) \in \mathcal{D}} \log \text{Pr} (i | u),
\end{equation}
where $\mathcal{D}$ is the collection of training data containing user-item interactions.
Since the number of items scales to billions, the sum operation of the denominator \eqref{next_item_prob} is computationally prohibitive.
Thus, we use the sampled softmax technique \cite{covington2016deep} to make the objective function trackable and choose the Adam optimizer \cite{kingma2014adam} for training MIND.

After training, the MIND network except for the label-aware attention layer can be used as user representation mapping function $\emph{f}_{user}$.
At serving time, user's behavior sequence and user profile are fed into the $\emph{f}_{user}$ function, producing multiple representation vectors for each user.
Then, these representation vectors are used to retrieve top $N$ items by an approximate nearest neighbor approach \cite{johnson2017billion}.
These items with highest similarities with user's representation vectors are retrieved and constitute the final set of candidate items for the matching stage of RS.
Please note that, when a user has new actions, it will alter his/her behavior sequence as well as the corresponding user representation vectors, thus MIND enables real-time personalization for the matching stage.

\subsection{Connections with Existing Methods}
Here, we make some remarks about the relations between MIND and two existing methods, illustrating their similarities as well as differences.

\textit{YouTube DNN.}
Both MIND and YouTube DNN utilize deep neural networks to model behavior data to generate user representations, which are used for large-scale item retrieval in the matching stage of industrial RS.
However, YouTube DNN uses one vector to represent a user while MIND uses multiple vectors for that.
When the value of $K$ in Algorithm \ref{alg:routing} equals to 1, MIND degenerates to YouTube DNN, thus MIND can be viewed as generalization of YouTube DNN.

\textit{DIN.}
In terms of capturing diverse interests of users, MIND and DIN share the similar goal.
However, the two methods differ in the way of achieving the goal as well as applicability.
To deal with diverse interests, DIN applies an attention mechanism at the item level, while MIND employs dynamic routing to generate interest capsules and considers diversity at the interest level.
Moreover, DIN focuses on the ranking stage as it handles thousands of items, however, MIND decouples the process of inferring user representations and measuring user-item compatibility, making it applicable to billion-scale items in the matching stage.

\section{Experiments}

\subsection{Offline Evaluation}
In this section, we present the comparisons between MIND and existing methods in terms of recommendation accuracy on several datasets under offline settings.

\subsubsection{Datasets and Experimental Setup}

\begin{table}[]
  \centering
  \caption{Statistics of the two datasets for offline evaluation.}
  \label{dataset:statistics}
  \begin{tabular}{ccccc}
    \toprule
    Dataset & Users & Goods & Categories & Samples \\ \hline
    Amazon Books & 351,356  & 393,801 & 1 & 6,271,511 \\ \hline
    TmallData & 2,014,865 & 934,751 & 6,377 & 50,929,802 \\ \bottomrule
  \end{tabular}
\end{table}

We choose two datasets for evaluating recommendation performance.
One is Amazon Books\footnote{http://jmcauley.ucsd.edu/data/amazon/} provided by \cite{he2016ups,mcauley2015image}, representing one of the most widely-used public dataset for e-commerce recommendations.
The other called TmallData is held out from Mobile Tmall App, containing historical behaviors of randomly sampled two millions of Tmall users in 10 days.
For Amazon Books, we only keep items which have been reviewed at least 10 times and users who have reviewed at least 10 items.
For TmallData, we filter out items clicked by less than 600 unique users.
The statistics of the two datasets are shown in Table~\ref{dataset:statistics}.

We choose next item prediction problem, that is predicting a user's next interaction, to evaluate the methods' performance, because it is the core task in the matching stage of RS.
After dividing the user-item interaction data of each dataset randomly into training set and test set by a ratio of 19:1, for each user, a randomly selected item interacted by the user is used as target item, while the items interacted before the target item are collected as the user behaviors.
Hit rate is adopted as the main metric to measure the recommendation performance, define as:
\begin{equation}
  HitRate@N = \frac{ \sum_{(u,i)\in \mathcal{D}_{test}} \emph{I}\,(\text{target item occurs in top } N) }{|\mathcal{D}_{test}|},
\end{equation}
where $\mathcal{D}_{test}$ denotes the test set consisting of pairs of users and target items $(u,i)$ and \emph{I} denotes the indicator function.

Hyperparameter tuning for the dimensionality of embedding vectors $d$ and the number of user interests $K$ is conducted by experiments on a group of parameters predefined according to the scale and distribution of each dataset, and each method is tested with best hyperparameters for a fair comparison.

\begin{table*}[]
  \centering
  \caption{HitRate of different methods on the two datasets, where best performance is in boldface.
HP denotes hyperparameters, including $K$ the number of interests and $d$ the dimensionality of embeddings. Only the results with hyperparameters having best performance is shown to demonstrate the effectiveness of corresponding methods. Percentages in the brackets indicate the relative improvements over YouTube DNN.}
  \label{dataset:hitrate}
  \begin{tabular}{c|c|c|ccccc}
    \toprule
    Dataset & HP & Metric & WALS & YouTube DNN & MaxMF-$K$-interest & MIND-1-interest & MIND-$K$-interest \\ \midrule
    \multirow{3}{*}{Amazon Books} & \multirow{3}{*}{\tabincell{c}{$K$ = 3 \\ $d$ = 36}} & HR@10 & 0.0144 (-37.66\%) & 0.0231 & 0.0285 (+23.38\%) & 0.0273 (+18.18\%) & \textbf{0.0309 (+33.77\%)} \\ \cline{3-8}
     & & HR@50 & 0.0553 (-25.87\%) & 0.0746 & 0.0862 (+15.55\%) & 0.0978 (+31.10\%) & \textbf{0.1101 (+47.59\%)} \\ \cline{3-8}
     & & HR@100 & 0.0907 (-20.65\%) & 0.1143 & 0.1304 (+14.09\%) & 0.1459 (+27.65\%) & \textbf{0.1631 (+42.69\%)} \\
    \midrule
    \multirow{3}{*}{TmallData} & \multirow{3}{*}{\tabincell{c}{$K$ = 5 \\ $d$ = 64}} & HR@10 & 0.0372 (-36.84\%) & 0.0589 & 0.0628 (+6.62\%) & 0.0720 (+22.24\%) & \textbf{0.0972 (+65.03\%)} \\ \cline{3-8}
     & & HR@50 & 0.0831 (-33.84\%) & 0.1256 & 0.1820 (+44.90\%) & 0.1512 (+20.38\%) & \textbf{0.2080 (+65.60\%)} \\ \cline{3-8}
     & & HR@100 & 0.1126 (-31.67\%) & 0.1648 & 0.2567 (+55.76\%) & 0.1930 (+17.11\%) & \textbf{0.2699 (+63.77\%)} \\
    \bottomrule
  \end{tabular}
\end{table*}

\subsubsection{Comparing Methods}
\begin{itemize}
  \item\textbf{WALS} \cite{aberger2016recommender}
  WALS, short for Weighted Alternating Least Squares, is a classical matrix factorization algorithm for decomposing user-item interaction matrix into hidden factors of users and items.
  Recommendation is made based on compatibilities between hidden factors of users and target items.
  \item\textbf{YouTube DNN} \cite{covington2016deep}
  As mentioned above, YouTube DNN is one of the most successful deep learning method used for industrial recommendation systems.
  \item\textbf{MaxMF} \cite{weston2013nonlinear}
  The method introduces a highly scalable method for learning nonlinear latent factorization to model multiple user interests.
\end{itemize}

\subsubsection{Experimental Results}

Table~\ref{dataset:hitrate} summarizes the performance of MIND as well as baselines on two datasets in terms of HitRate@N (N = 10, 50, 100).
Clearly, MIND accomplishes comparable performance to all of the baselines on both datasets.
The matrix factorization approach, WALS, is beaten by other methods, revealing the power of deep learning for improving the matching stage of RS.
However, equipped without deep learning, MaxMF performs much better than WALS, which can be explained by the fact that MaxMF generalizes standard MF to a nonlinear model and adopts multiple user representation vectors.
It can be observed that methods employing multiple user representation vectors (MaxMF-$K$-interest, MIND-$K$-interest) performs generally better than other methods (WALS, YouTube DNN, MIND-1-interest).
Therefore, using multiple user representation vectors is proved to be an effective way for modeling user's diverse interests as well as boosting recommendation accuracy.
Moreover, we can observe that the improvement introduced by multiple user representation vectors is more significant for TmallData, as the users of Tmall tend to exhibit more diverse interests.
This increasement of diversity can also be reflected by the best $K$ for each dataset, where
the best $K$ for TmallData is larger than that for Amazon Books.
The improvement of MIND-1-interest over YouTube DNN shows that dynamic routing serves as a better pooling strategy than average pooling.
Considering the results of MaxMF and MIND-$K$-interest, it verifies that extracting multiple interests from user behaviors by dynamic routing outperforms the nonlinear modeling strategy used in MaxMF.
This can be attributed to two points:
1) The multi-interest extractor layer utilizes a clustering procedure for generating interest representations, which achieves more precise representation of user.
2) Label-aware attention layer makes target item attend over multiple user representation vectors, enabling more accurate matching between user interests and target item.

\begin{figure}[h]
  \centering
  \includegraphics[scale=0.4]{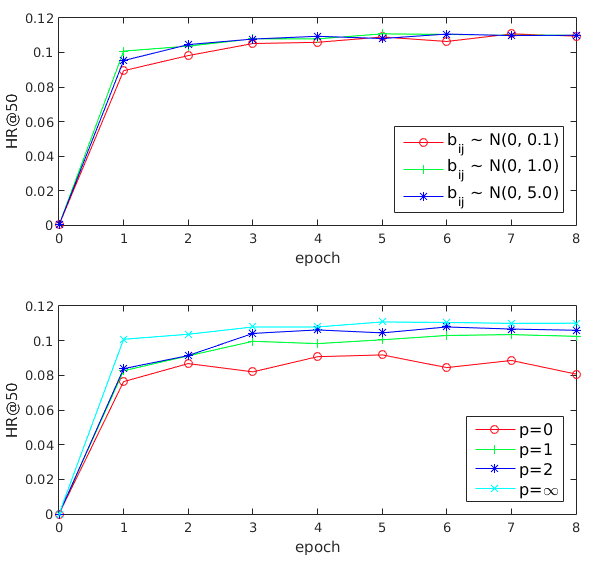}
  \caption{Hyperparameters' impact. The upper part shows that MIND can achieve comparable results with different $\sigma$; the lower part shows that MIND performs better with bigger $p$.}
  \label{fig:analysis}
\end{figure}

\subsection{Analysis of Hyperparameters}

In this section, we conduct two experiments on Amazon Books to study the influence of the hyperparameters within multi-interest extractor layer and label-aware attention layer.

\textbf{Initialization of routing logits.}
The random initialization for routing logits adopted in multi-interest extractor layer is similar to the initialization of K-means centroids, where the distributions of initial cluster centers have strong impact on the final clustering results.
As the routing logits are initialized according to gaussian distribution $\mathcal N(0, \sigma ^ 2)$, we concern about different values of $\sigma$ may lead to different convergence which has effect on the performance.
To study the impact of $\sigma$, we initialize the routing logits $b_{ij}$ with 3 different values of $\sigma$, $0.1$, $1$ and $5$.
The results are shown by the upper part of Figure \ref{fig:analysis}, where each curve of 3 values almost overlap.
This observation reveals that MIND is robust to the values of $\sigma$, and it is rational to choose $\sigma = 1$ for our practical applications.

\textbf{Power number in label-aware attention.}
As mentioned before, the power number $p$ within label-aware attention controls the proportion of each interest to the combined label-aware interest representation.
We compare the performance of MIND as $p$ varies from $0$ to $\infty$ and show the results by the lower part of Figure \ref{fig:analysis}.
Clearly, the performance of $p = 0$ is much worse than the others.
The reason is that, when taking $p = 0$ each interest has the same attention thus the combined interest representation equals the average of interests with no reference to the label.
Taking $p \geqslant 1$, the attention scores are proportional to the similarities between interest representation vectors and target item embeddings, which makes the combined interest representation a weighted sum of interests.
It also shows that performance gets better as $p$ increases, since the representation vector of the interest with more similarity to the target item acquires larger attention, which evolves to a hard attention scheme as $p = \infty$.
By this scheme, the interest representation nearest to the target item dominates the combined interest representation, enabling MIND converge faster and perform the best.

\subsection{Online Experiments}

\begin{figure}[h]
  \centering
  \includegraphics[scale=0.22]{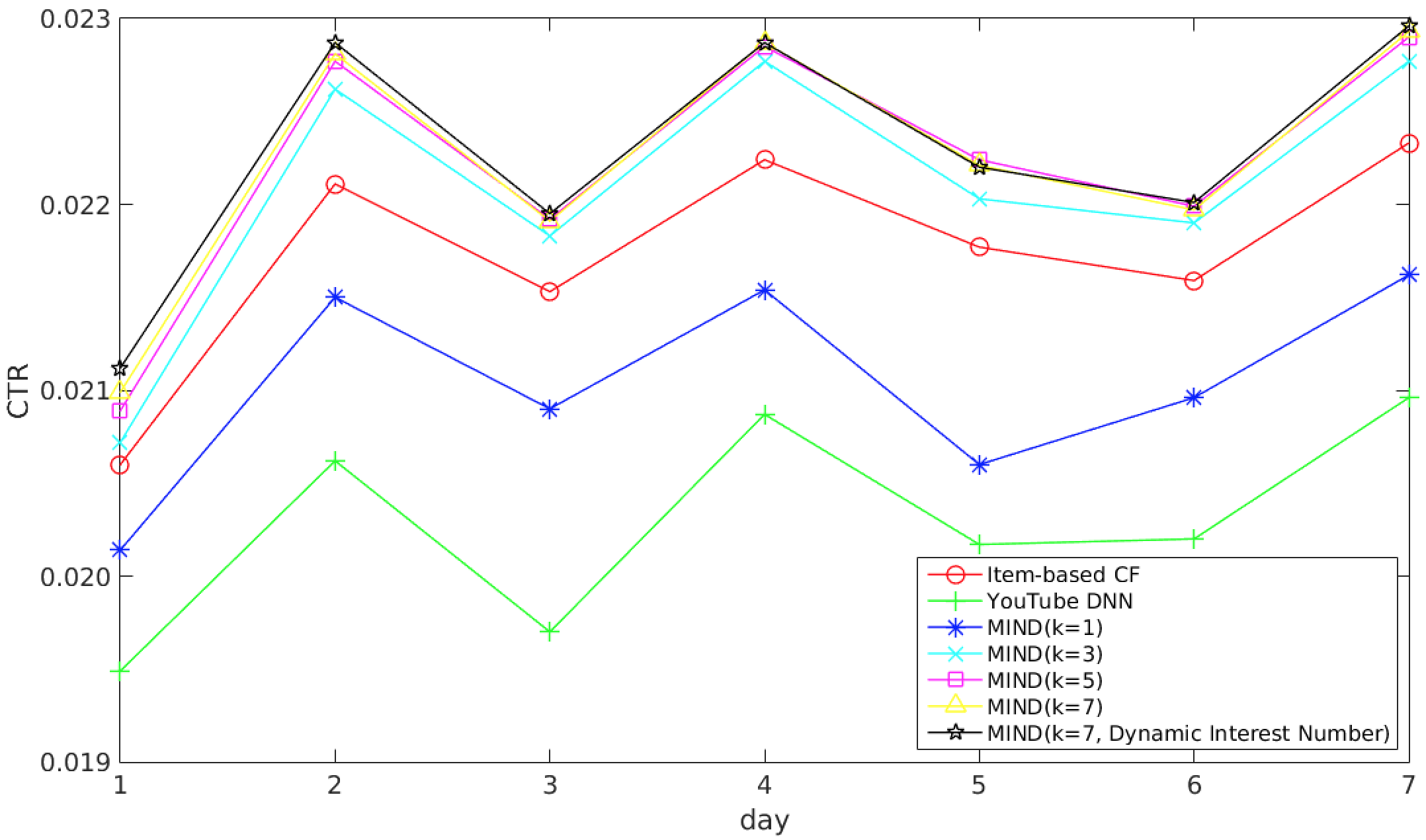}
  \caption{Online CTRs in a week. MIND with 5 \textasciitilde 7 interests performs best in all comparing methods. MIND significantly beats the two baseline methods, item-based CF and YouTube DNN.}
  \label{fig:online_results}
\end{figure}

We conduct online experiments by deploying MIND to handle real traffic at Tmall homepage for one week.
To make comparisons fairly, all methods deployed in the matching stage are followed by the same ranking procedure.
CTR, short for click-through-rate, a widely used industrial metric, is used to measure the performance of methods for serving online traffic.

There are two baseline methods for online experiments.
One is item-based CF, which is the base matching algorithm serving the majority of the online traffic.
The other is YouTube DNN, which is the well-known deep learning-based matching model.
We deploy all comparing methods in an A/B test framework, and one thousand of candidate items are retrieved by each method, which then fed to the ranking stage for final recommendation.

The experimental results are summarized in Figure \ref{fig:online_results}.
It is clearly that MIND outperforms item-based CF and YouTube DNN, which indicates that MIND generates a better user representation.
Besides, we make the following observations:
1) As is optimized by the long-term practice, item-based CF performs better than YouTube DNN which is also exceeded by MIND with single interest.
2) A very noticeable trend is that the performance of MIND gets better as the number of extracted interests increases from 1 to 5.
3) The performance of MIND peaks when the number of extracted interests reaches 5, after that the CTR remains constant and the improvement of 7 interests is ignorable.
4) MIND with dynamic interest number has the comparable performance with MIND with 7 interests.
From the observations above, we make several conclusions.
First, for Tmall, the optimal number of user interests is 5 \textasciitilde 7, which reveals the average diversity of user interests.
Second, the dynamic interest number mechanism does not bring CTR gain, but during the experiments we recognize the scheme can decrease the cost of serving, which benefits large-scale service such as Tmall and is more adoptable in practice.
In a word, the online experiments validate that MIND achieves an better solution to model users with diverse interests and can significantly advance the whole RS.

\subsection{Case Study}

\subsubsection{Coupling Coefficients.}

\begin{figure}[ht]
  \centering
  \includegraphics[scale=0.4]{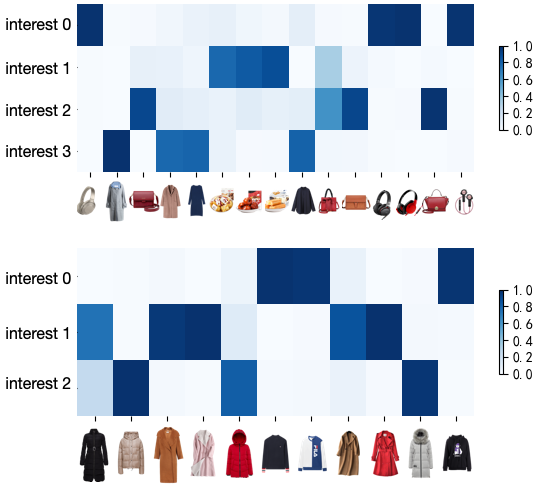}
  \caption{Heatmap of coupling coefficients for two users. Each class of behaviors has the max coupling coefficients on the corresponding interest. User C (upper) and user D (below) have different granularity of interests.}
  \label{fig:heatmap}
\end{figure}

The coupling coefficients between behavior capsules and interest capsules quantify the grade of membership of behaviors to interests.
In this section, we visualize these coupling coefficients to show that the interest extraction process is interpretable.

Figure \ref{fig:heatmap} illustrates the coupling coefficients associated to two users randomly selected from Tmall daily active users, where each row corresponds to one interest capsule and each column corresponds to one behavior.
It shows that user C (upper) has interacted with 4 classes of goods (headphones, snacks, handbags and clothes), each of which has the max coupling coefficients on one interest capsule and forms the corresponding interest.
While user D (below) is interested only in clothes, thus the 3 interests with finer grain size (sweaters, overcoats and down jackets) are resolved from the behaviors.
Regarding this result, we confirm that each class of user behaviors are clustered together and form the corresponding interest representation vector.

\subsubsection{Item Distribution.}

\begin{figure}[ht]
  \centering
  \includegraphics[scale=0.3]{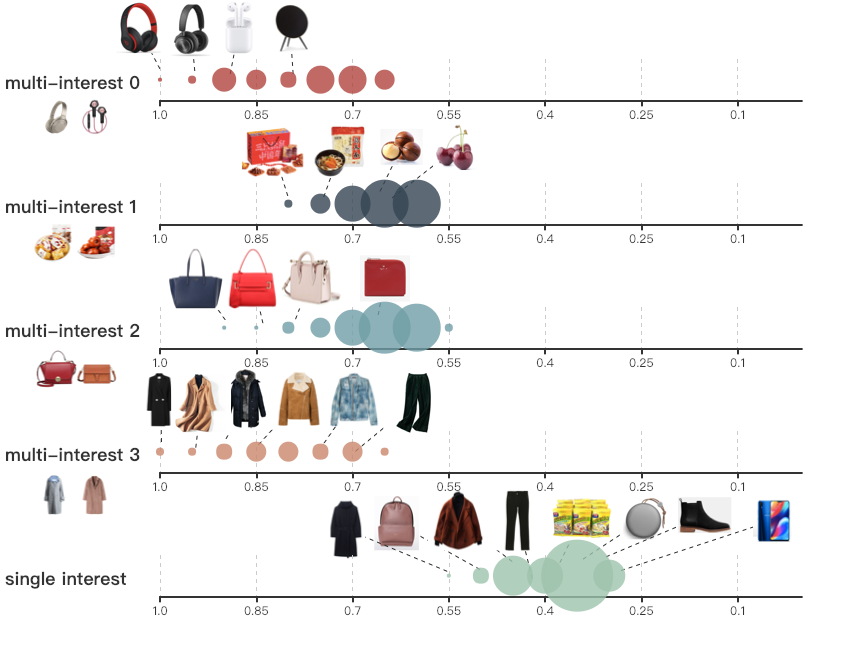}
  \caption{The distribution of items recalled by each interest corresponding to the user behaviors exampled on the left. Each interest is demonstrated by one axis, of which the coordinate is the similarity between items and interests. The size of the point is proportional to the number of the items with the specific similarity.}
  \label{fig:distribution}
\end{figure}

At serving time, items similar to user interests are retrieved by nearest neighbor search.
We visualize the distribution of these items recalled by each interest based on their similarity to the corresponding interest.
Figure \ref{fig:distribution} shows the item distributions of the same user (user C) mentioned by Figure \ref{fig:heatmap} (upper).
The distributions are obtained by two methods respectively, where the upper 4 axes demonstrate the items recalled by 4 interests based on MIND while the lowest axis illustrates that based on YouTube DNN.
The items are scattered at the axes according to their similarity with the interests, which has been scaled to 0 \textasciitilde 1 by min-max normalization and rounded to the nearest 0.5.
One point is assembled by the items lying within the specific range, thus the size of each point represents the number of the items with the corresponding similarity.
We also show some items selected randomly from all the candidates.
As expected, the items recalled by MIND are strongly correlated with the corresponding interest, while that by YouTube DNN vary widely along the categories of items and have lower similarity to the user's behaviors.

\section{System Deployment}

In this section, we describe the implementation and deployment of MIND at Tmall.
A typical workflow composed of several basic platforms is shown as Figure \ref{fig:deployment} and detailed as below:

\begin{figure}[ht]
  \centering
  \includegraphics[scale=0.35]{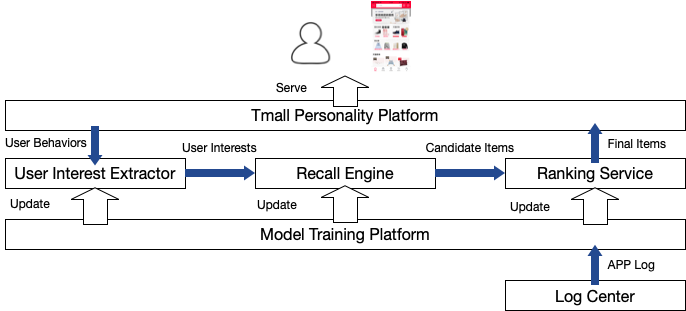}
  \caption{Architecture of the RS at Tmall.}
  \label{fig:deployment}
\end{figure}

As users launch Mobile Tmall APP, the requests for recommendation are sent to \textit{Tmall Personality Platform}, the server cluster integrated with a bunch of plug-in modules and served as online recommender service of Tmall.
Recent behaviors of users are retrieved by \textit{Tmall Personality Platform} and sent to \textit{User Interest Extractor} which is the main module implementing MIND for transforming user behaviors to multiple user interests.
Subsequently, \textit{Recall Engine} searches for the items with embedding vectors nearest to the user interests.
Items triggered by different interests are merged together as candidate items and sorted by their similarity to the user interests.
The whole procedure of selecting thousands of candidate items from the billion-scale item pool by \textit{User Interest Extractor} and \textit{Recall Engine} can be fulfilled in less than 15 milliseconds, due to the effectiveness of serving based on MIND.
Taking a tradeoff between the scope of items and the response time of the system, top 1000 of these candidate items are scored by \textit{Ranking Service} which predicts CTRs with a bunch of features.
Finally, Tmall Personality Platform completes the item list as the recommendation results shown to users.
Both \textit{User Interest Extractor} and \textit{Ranking Service}
are trained on \textit{Model Training Platform} using 100 GPUs, by which the training can be executed in 8 hours.
Benefiting from the superior performance of \textit{Model Training Platform}, the deep network served for prediction is updated every day, which guarantees the newly released products to be calculated and exposured.

\section{Conclusion and Future Work}
In this paper, we propose a new structure of neural network, namely Multi-Interest Network with Dynamic routing (MIND), to represent user's diverse interests for the matching stage in e-commerce recommendation, which involves billion scale users and items.
Specifically, we design a multi-interest extractor layer with a variant dynamic routing to extract user's diverse interests and these interests are then trained with a novel label-aware attention scheme.
Offline experiments are conducted to demonstrate that MIND achieves superior performance on public benchmarks.
Online CTRs are also reported to demonstrate the effectiveness and feasibility of MIND at Tmall's live production.
For future work, we will pursue two directions.
The first is to incorporate more information about user's behavior sequence, such as behavior time etc.
The second direction is to optimize the initialization scheme of dynamic routing, referring to K-means++ initialization scheme, so as to achieve a better user representation.

%
% The acknowledgments section is defined using the "acks" environment (and NOT an unnumbered section). This ensures
% the proper identification of the section in the article metadata, and the consistent spelling of the heading.
\begin{acks}
We would like to thank colleagues of our team - Jizhe Wang, Andreas Pfadler, Jiaming Xu, Wen Chen, Lifeng Wang, Xin Guo and Cheng Guo for useful discussions and supports on this work.
We are grateful to our cooperative team - search engineering team.
We also thank the anonymous reviewers for their valuable comments and suggestions that help improve the quality of this manuscript.
\end{acks}

%
% The next two lines define the bibliography style to be used, and the bibliography file.
\bibliographystyle{ACM-Reference-Format}
\bibliography{umi.bib}


\begin{thebibliography}{31}

%%% the Web address.

\ifx \showCODEN    \undefined \def \showCODEN     #1{\unskip}     \fi
\ifx \showDOI      \undefined \def \showDOI       #1{#1}\fi
\ifx \showISBNx    \undefined \def \showISBNx     #1{\unskip}     \fi
\ifx \showISBNxiii \undefined \def \showISBNxiii  #1{\unskip}     \fi
\ifx \showISSN     \undefined \def \showISSN      #1{\unskip}     \fi
\ifx \showLCCN     \undefined \def \showLCCN      #1{\unskip}     \fi
\ifx \shownote     \undefined \def \shownote      #1{#1}          \fi
\ifx \showarticletitle \undefined \def \showarticletitle #1{#1}   \fi
\ifx \showURL      \undefined \def \showURL       {\relax}        \fi
% invisible to TeX
\providecommand\bibfield[2]{#2}
\providecommand\bibinfo[2]{#2}
\providecommand\natexlab[1]{#1}
\providecommand\showeprint[2][]{arXiv:#2}

\bibitem[\protect\citeauthoryear{Aberger}{Aberger}{2016}]%
        {aberger2016recommender}
\bibfield{author}{\bibinfo{person}{Christopher~R Aberger}.}
  \bibinfo{year}{2016}\natexlab{}.
\newblock \bibinfo{booktitle}{\emph{Recommender: An analysis of collaborative
  filtering techniques}}.
\newblock \bibinfo{type}{{T}echnical {R}eport}.
\newblock


\bibitem[\protect\citeauthoryear{Amir, Wallace, Lyu, Carvalho, and Silva}{Amir
  et~al\mbox{.}}{2016}]%
        {amir2016modelling}
\bibfield{author}{\bibinfo{person}{Silvio Amir}, \bibinfo{person}{Byron~C.
  Wallace}, \bibinfo{person}{Hao Lyu}, \bibinfo{person}{Paula Carvalho}, {and}
  \bibinfo{person}{Mario~J. Silva}.} \bibinfo{year}{2016}\natexlab{}.
\newblock \showarticletitle{Modelling Context with User Embeddings for Sarcasm
  Detection in Social Media}. In \bibinfo{booktitle}{\emph{Proceedings of The
  20th SIGNLL Conference on Computational Natural Language Learning}}.
  \bibinfo{publisher}{Association for Computational Linguistics},
  \bibinfo{pages}{167--177}.
\newblock


\bibitem[\protect\citeauthoryear{Batmaz, Yurekli, Bilge, and Kaleli}{Batmaz
  et~al\mbox{.}}{2018}]%
        {batmaz2018review}
\bibfield{author}{\bibinfo{person}{Zeynep Batmaz}, \bibinfo{person}{Ali
  Yurekli}, \bibinfo{person}{Alper Bilge}, {and} \bibinfo{person}{Cihan
  Kaleli}.} \bibinfo{year}{2018}\natexlab{}.
\newblock \showarticletitle{A review on deep learning for recommender systems:
  challenges and remedies}.
\newblock \bibinfo{journal}{\emph{Artificial Intelligence Review}}
  (\bibinfo{year}{2018}), \bibinfo{pages}{1--37}.
\newblock


\bibitem[\protect\citeauthoryear{Bell and Koren}{Bell and Koren}{2007}]%
        {bell2007improved}
\bibfield{author}{\bibinfo{person}{Robert~M Bell} {and} \bibinfo{person}{Yehuda
  Koren}.} \bibinfo{year}{2007}\natexlab{}.
\newblock \showarticletitle{Improved neighborhood-based collaborative
  filtering}. In \bibinfo{booktitle}{\emph{KDD cup and workshop at the 13th ACM
  SIGKDD international conference on knowledge discovery and data mining}}.
  Citeseer, \bibinfo{pages}{7--14}.
\newblock


\bibitem[\protect\citeauthoryear{Cantador, Bellog{\'\i}n, and Vallet}{Cantador
  et~al\mbox{.}}{2010}]%
        {cantador2010content}
\bibfield{author}{\bibinfo{person}{Iv{\'a}n Cantador},
  \bibinfo{person}{Alejandro Bellog{\'\i}n}, {and} \bibinfo{person}{David
  Vallet}.} \bibinfo{year}{2010}\natexlab{}.
\newblock \showarticletitle{Content-based recommendation in social tagging
  systems}. In \bibinfo{booktitle}{\emph{Proceedings of the fourth ACM
  conference on Recommender systems}}. ACM, \bibinfo{pages}{237--240}.
\newblock


\bibitem[\protect\citeauthoryear{Chen, Xu, He, Xia, and Wang}{Chen
  et~al\mbox{.}}{2016}]%
        {chen2016learning}
\bibfield{author}{\bibinfo{person}{Tao Chen}, \bibinfo{person}{Ruifeng Xu},
  \bibinfo{person}{Yulan He}, \bibinfo{person}{Yunqing Xia}, {and}
  \bibinfo{person}{Xuan Wang}.} \bibinfo{year}{2016}\natexlab{}.
\newblock \showarticletitle{Learning user and product distributed
  representations using a sequence model for sentiment analysis}.
\newblock \bibinfo{journal}{\emph{IEEE Computational Intelligence Magazine}}
  \bibinfo{volume}{11}, \bibinfo{number}{3} (\bibinfo{year}{2016}),
  \bibinfo{pages}{34--44}.
\newblock


\bibitem[\protect\citeauthoryear{Covington, Adams, and Sargin}{Covington
  et~al\mbox{.}}{2016}]%
        {covington2016deep}
\bibfield{author}{\bibinfo{person}{Paul Covington}, \bibinfo{person}{Jay
  Adams}, {and} \bibinfo{person}{Emre Sargin}.}
  \bibinfo{year}{2016}\natexlab{}.
\newblock \showarticletitle{Deep neural networks for youtube recommendations}.
  In \bibinfo{booktitle}{\emph{Proceedings of the 10th ACM Conference on
  Recommender Systems}}. ACM, \bibinfo{pages}{191--198}.
\newblock


\bibitem[\protect\citeauthoryear{Elkahky, Song, and He}{Elkahky
  et~al\mbox{.}}{2015}]%
        {elkahky2015multi}
\bibfield{author}{\bibinfo{person}{Ali~Mamdouh Elkahky}, \bibinfo{person}{Yang
  Song}, {and} \bibinfo{person}{Xiaodong He}.} \bibinfo{year}{2015}\natexlab{}.
\newblock \showarticletitle{A multi-view deep learning approach for cross
  domain user modeling in recommendation systems}. In
  \bibinfo{booktitle}{\emph{Proceedings of the 24th International Conference on
  World Wide Web}}. International World Wide Web Conferences Steering
  Committee, \bibinfo{pages}{278--288}.
\newblock


\bibitem[\protect\citeauthoryear{Guo, Tang, Ye, Li, and He}{Guo
  et~al\mbox{.}}{2017}]%
        {guo2017deepfm}
\bibfield{author}{\bibinfo{person}{Huifeng Guo}, \bibinfo{person}{Ruiming
  Tang}, \bibinfo{person}{Yunming Ye}, \bibinfo{person}{Zhenguo Li}, {and}
  \bibinfo{person}{Xiuqiang He}.} \bibinfo{year}{2017}\natexlab{}.
\newblock \showarticletitle{DeepFM: A Factorization-machine Based Neural
  Network for CTR Prediction}. In \bibinfo{booktitle}{\emph{Proceedings of the
  26th International Joint Conference on Artificial Intelligence}}
  \emph{(\bibinfo{series}{IJCAI'17})}. \bibinfo{publisher}{AAAI Press},
  \bibinfo{pages}{1725--1731}.
\newblock
\showISBNx{978-0-9992411-0-3}


\bibitem[\protect\citeauthoryear{He and McAuley}{He and McAuley}{2016}]%
        {he2016ups}
\bibfield{author}{\bibinfo{person}{Ruining He} {and} \bibinfo{person}{Julian
  McAuley}.} \bibinfo{year}{2016}\natexlab{}.
\newblock \showarticletitle{Ups and downs: Modeling the visual evolution of
  fashion trends with one-class collaborative filtering}. In
  \bibinfo{booktitle}{\emph{proceedings of the 25th international conference on
  world wide web}}. International World Wide Web Conferences Steering
  Committee, \bibinfo{pages}{507--517}.
\newblock


\bibitem[\protect\citeauthoryear{He, Liao, Zhang, Nie, Hu, and Chua}{He
  et~al\mbox{.}}{2017}]%
        {he2017neural}
\bibfield{author}{\bibinfo{person}{Xiangnan He}, \bibinfo{person}{Lizi Liao},
  \bibinfo{person}{Hanwang Zhang}, \bibinfo{person}{Liqiang Nie},
  \bibinfo{person}{Xia Hu}, {and} \bibinfo{person}{Tat-Seng Chua}.}
  \bibinfo{year}{2017}\natexlab{}.
\newblock \showarticletitle{Neural collaborative filtering}. In
  \bibinfo{booktitle}{\emph{Proceedings of the 26th International Conference on
  World Wide Web}}. International World Wide Web Conferences Steering
  Committee, \bibinfo{pages}{173--182}.
\newblock


\bibitem[\protect\citeauthoryear{Herlocker, Konstan, and Riedl}{Herlocker
  et~al\mbox{.}}{2002}]%
        {herlocker2002empirical}
\bibfield{author}{\bibinfo{person}{Jon Herlocker}, \bibinfo{person}{Joseph~A
  Konstan}, {and} \bibinfo{person}{John Riedl}.}
  \bibinfo{year}{2002}\natexlab{}.
\newblock \showarticletitle{An empirical analysis of design choices in
  neighborhood-based collaborative filtering algorithms}.
\newblock \bibinfo{journal}{\emph{Information retrieval}} \bibinfo{volume}{5},
  \bibinfo{number}{4} (\bibinfo{year}{2002}), \bibinfo{pages}{287--310}.
\newblock


\bibitem[\protect\citeauthoryear{Hinton, Krizhevsky, and Wang}{Hinton
  et~al\mbox{.}}{2011}]%
        {hinton2011transforming}
\bibfield{author}{\bibinfo{person}{Geoffrey~E Hinton}, \bibinfo{person}{Alex
  Krizhevsky}, {and} \bibinfo{person}{Sida~D Wang}.}
  \bibinfo{year}{2011}\natexlab{}.
\newblock \showarticletitle{Transforming auto-encoders}. In
  \bibinfo{booktitle}{\emph{International Conference on Artificial Neural
  Networks}}. Springer, \bibinfo{pages}{44--51}.
\newblock


\bibitem[\protect\citeauthoryear{Hinton, Sabour, and Frosst}{Hinton
  et~al\mbox{.}}{2018}]%
        {hinton2018matrix}
\bibfield{author}{\bibinfo{person}{Geoffrey~E Hinton}, \bibinfo{person}{Sara
  Sabour}, {and} \bibinfo{person}{Nicholas Frosst}.}
  \bibinfo{year}{2018}\natexlab{}.
\newblock \showarticletitle{Matrix capsules with {EM} routing}. In
  \bibinfo{booktitle}{\emph{International Conference on Learning
  Representations}}.
\newblock


\bibitem[\protect\citeauthoryear{Johnson, Douze, and J{\'e}gou}{Johnson
  et~al\mbox{.}}{2017}]%
        {johnson2017billion}
\bibfield{author}{\bibinfo{person}{Jeff Johnson}, \bibinfo{person}{Matthijs
  Douze}, {and} \bibinfo{person}{Herv{\'e} J{\'e}gou}.}
  \bibinfo{year}{2017}\natexlab{}.
\newblock \showarticletitle{Billion-scale similarity search with gpus}.
\newblock \bibinfo{journal}{\emph{arXiv preprint arXiv:1702.08734}}
  (\bibinfo{year}{2017}).
\newblock


\bibitem[\protect\citeauthoryear{Kingma and Ba}{Kingma and Ba}{2014}]%
        {kingma2014adam}
\bibfield{author}{\bibinfo{person}{Diederik~P Kingma} {and}
  \bibinfo{person}{Jimmy Ba}.} \bibinfo{year}{2014}\natexlab{}.
\newblock \showarticletitle{Adam: A method for stochastic optimization}.
\newblock \bibinfo{journal}{\emph{arXiv preprint arXiv:1412.6980}}
  (\bibinfo{year}{2014}).
\newblock


\bibitem[\protect\citeauthoryear{Koren, Bell, and Volinsky}{Koren
  et~al\mbox{.}}{2009}]%
        {koren2009matrix}
\bibfield{author}{\bibinfo{person}{Yehuda Koren}, \bibinfo{person}{Robert
  Bell}, {and} \bibinfo{person}{Chris Volinsky}.}
  \bibinfo{year}{2009}\natexlab{}.
\newblock \showarticletitle{Matrix factorization techniques for recommender
  systems}.
\newblock \bibinfo{journal}{\emph{Computer}} \bibinfo{number}{8}
  (\bibinfo{year}{2009}), \bibinfo{pages}{30--37}.
\newblock


\bibitem[\protect\citeauthoryear{LaLonde and Bagci}{LaLonde and Bagci}{2018}]%
        {lalonde2018capsules}
\bibfield{author}{\bibinfo{person}{Rodney LaLonde} {and} \bibinfo{person}{Ulas
  Bagci}.} \bibinfo{year}{2018}\natexlab{}.
\newblock \showarticletitle{Capsules for Object Segmentation}.
\newblock \bibinfo{journal}{\emph{arXiv preprint arXiv:1804.04241}}
  (\bibinfo{year}{2018}).
\newblock


\bibitem[\protect\citeauthoryear{LeCun, Bengio, and Hinton}{LeCun
  et~al\mbox{.}}{2015}]%
        {lecun2015deep}
\bibfield{author}{\bibinfo{person}{Yann LeCun}, \bibinfo{person}{Yoshua
  Bengio}, {and} \bibinfo{person}{Geoffrey Hinton}.}
  \bibinfo{year}{2015}\natexlab{}.
\newblock \showarticletitle{Deep learning}.
\newblock \bibinfo{journal}{\emph{nature}} \bibinfo{volume}{521},
  \bibinfo{number}{7553} (\bibinfo{year}{2015}), \bibinfo{pages}{436}.
\newblock


\bibitem[\protect\citeauthoryear{McAuley, Targett, Shi, and Van
  Den~Hengel}{McAuley et~al\mbox{.}}{2015}]%
        {mcauley2015image}
\bibfield{author}{\bibinfo{person}{Julian McAuley},
  \bibinfo{person}{Christopher Targett}, \bibinfo{person}{Qinfeng Shi}, {and}
  \bibinfo{person}{Anton Van Den~Hengel}.} \bibinfo{year}{2015}\natexlab{}.
\newblock \showarticletitle{Image-based recommendations on styles and
  substitutes}. In \bibinfo{booktitle}{\emph{Proceedings of the 38th
  International ACM SIGIR Conference on Research and Development in Information
  Retrieval}}. ACM, \bibinfo{pages}{43--52}.
\newblock


\bibitem[\protect\citeauthoryear{Sabour, Frosst, and Hinton}{Sabour
  et~al\mbox{.}}{2017}]%
        {sabour2017dynamic}
\bibfield{author}{\bibinfo{person}{Sara Sabour}, \bibinfo{person}{Nicholas
  Frosst}, {and} \bibinfo{person}{Geoffrey~E Hinton}.}
  \bibinfo{year}{2017}\natexlab{}.
\newblock \showarticletitle{Dynamic routing between capsules}. In
  \bibinfo{booktitle}{\emph{Advances in Neural Information Processing
  Systems}}. \bibinfo{pages}{3856--3866}.
\newblock


\bibitem[\protect\citeauthoryear{Sarwar, Karypis, Konstan, and Riedl}{Sarwar
  et~al\mbox{.}}{2001}]%
        {sarwar2001item}
\bibfield{author}{\bibinfo{person}{Badrul Sarwar}, \bibinfo{person}{George
  Karypis}, \bibinfo{person}{Joseph Konstan}, {and} \bibinfo{person}{John
  Riedl}.} \bibinfo{year}{2001}\natexlab{}.
\newblock \showarticletitle{Item-based collaborative filtering recommendation
  algorithms}. In \bibinfo{booktitle}{\emph{Proceedings of the 10th
  international conference on World Wide Web}}. ACM, \bibinfo{pages}{285--295}.
\newblock


\bibitem[\protect\citeauthoryear{Tang and Wang}{Tang and Wang}{2018}]%
        {tang2018personalized}
\bibfield{author}{\bibinfo{person}{Jiaxi Tang} {and} \bibinfo{person}{Ke
  Wang}.} \bibinfo{year}{2018}\natexlab{}.
\newblock \showarticletitle{Personalized top-n sequential recommendation via
  convolutional sequence embedding}. In \bibinfo{booktitle}{\emph{Proceedings
  of the Eleventh ACM International Conference on Web Search and Data Mining}}.
  ACM, \bibinfo{pages}{565--573}.
\newblock


\bibitem[\protect\citeauthoryear{Vaswani, Shazeer, Parmar, Uszkoreit, Jones,
  Gomez, Kaiser, and Polosukhin}{Vaswani et~al\mbox{.}}{2017}]%
        {vaswani2017attention}
\bibfield{author}{\bibinfo{person}{Ashish Vaswani}, \bibinfo{person}{Noam
  Shazeer}, \bibinfo{person}{Niki Parmar}, \bibinfo{person}{Jakob Uszkoreit},
  \bibinfo{person}{Llion Jones}, \bibinfo{person}{Aidan~N Gomez},
  \bibinfo{person}{{\L}ukasz Kaiser}, {and} \bibinfo{person}{Illia
  Polosukhin}.} \bibinfo{year}{2017}\natexlab{}.
\newblock \showarticletitle{Attention is all you need}. In
  \bibinfo{booktitle}{\emph{Advances in Neural Information Processing
  Systems}}. \bibinfo{pages}{5998--6008}.
\newblock


\bibitem[\protect\citeauthoryear{Wang, Huang, Zhao, Zhang, Zhao, and Lee}{Wang
  et~al\mbox{.}}{2018}]%
        {wang2018billion}
\bibfield{author}{\bibinfo{person}{Jizhe Wang}, \bibinfo{person}{Pipei Huang},
  \bibinfo{person}{Huan Zhao}, \bibinfo{person}{Zhibo Zhang},
  \bibinfo{person}{Binqiang Zhao}, {and} \bibinfo{person}{Dik~Lun Lee}.}
  \bibinfo{year}{2018}\natexlab{}.
\newblock \showarticletitle{Billion-scale Commodity Embedding for E-commerce
  Recommendation in Alibaba}. In \bibinfo{booktitle}{\emph{Proceedings of the
  24th ACM SIGKDD International Conference on Knowledge Discovery \&\#38; Data
  Mining}} \emph{(\bibinfo{series}{KDD '18})}. \bibinfo{pages}{839--848}.
\newblock
\showISBNx{978-1-4503-5552-0}


\bibitem[\protect\citeauthoryear{Weston, Weiss, and Yee}{Weston
  et~al\mbox{.}}{2013}]%
        {weston2013nonlinear}
\bibfield{author}{\bibinfo{person}{Jason Weston}, \bibinfo{person}{Ron~J
  Weiss}, {and} \bibinfo{person}{Hector Yee}.} \bibinfo{year}{2013}\natexlab{}.
\newblock \showarticletitle{Nonlinear latent factorization by embedding
  multiple user interests}. In \bibinfo{booktitle}{\emph{Proceedings of the 7th
  ACM conference on Recommender systems}}. ACM, \bibinfo{pages}{65--68}.
\newblock


\bibitem[\protect\citeauthoryear{Xue, Dai, Zhang, Huang, and Chen}{Xue
  et~al\mbox{.}}{2017}]%
        {xue2017deep}
\bibfield{author}{\bibinfo{person}{Hong-Jian Xue}, \bibinfo{person}{Xinyu Dai},
  \bibinfo{person}{Jianbing Zhang}, \bibinfo{person}{Shujian Huang}, {and}
  \bibinfo{person}{Jiajun Chen}.} \bibinfo{year}{2017}\natexlab{}.
\newblock \showarticletitle{Deep Matrix Factorization Models for Recommender
  Systems.}. In \bibinfo{booktitle}{\emph{IJCAI}}. \bibinfo{pages}{3203--3209}.
\newblock


\bibitem[\protect\citeauthoryear{Yang, Zhao, Ye, Lei, Zhao, and Zhang}{Yang
  et~al\mbox{.}}{2018}]%
        {zhao2018investigating}
\bibfield{author}{\bibinfo{person}{Min Yang}, \bibinfo{person}{Wei Zhao},
  \bibinfo{person}{Jianbo Ye}, \bibinfo{person}{Zeyang Lei},
  \bibinfo{person}{Zhou Zhao}, {and} \bibinfo{person}{Soufei Zhang}.}
  \bibinfo{year}{2018}\natexlab{}.
\newblock \showarticletitle{Investigating Capsule Networks with Dynamic Routing
  for Text Classification}. In \bibinfo{booktitle}{\emph{Proceedings of the
  2018 Conference on Empirical Methods in Natural Language Processing}}.
  \bibinfo{publisher}{Association for Computational Linguistics},
  \bibinfo{pages}{3110--3119}.
\newblock


\bibitem[\protect\citeauthoryear{Yin, Cui, Chen, Hu, and Zhou}{Yin
  et~al\mbox{.}}{2015}]%
        {yin2015dynamic}
\bibfield{author}{\bibinfo{person}{Hongzhi Yin}, \bibinfo{person}{Bin Cui},
  \bibinfo{person}{Ling Chen}, \bibinfo{person}{Zhiting Hu}, {and}
  \bibinfo{person}{Xiaofang Zhou}.} \bibinfo{year}{2015}\natexlab{}.
\newblock \showarticletitle{Dynamic user modeling in social media systems}.
\newblock \bibinfo{journal}{\emph{ACM Transactions on Information Systems
  (TOIS)}} \bibinfo{volume}{33}, \bibinfo{number}{3} (\bibinfo{year}{2015}),
  \bibinfo{pages}{10}.
\newblock


\bibitem[\protect\citeauthoryear{Yu, Wan, and Zhou}{Yu et~al\mbox{.}}{2016}]%
        {yu2016user}
\bibfield{author}{\bibinfo{person}{Yang Yu}, \bibinfo{person}{Xiaojun Wan},
  {and} \bibinfo{person}{Xinjie Zhou}.} \bibinfo{year}{2016}\natexlab{}.
\newblock \showarticletitle{User embedding for scholarly microblog
  recommendation}. In \bibinfo{booktitle}{\emph{Proceedings of the 54th Annual
  Meeting of the Association for Computational Linguistics (Volume 2: Short
  Papers)}}, Vol.~\bibinfo{volume}{2}. \bibinfo{pages}{449--453}.
\newblock


\bibitem[\protect\citeauthoryear{Zhou, Zhu, Song, Fan, Zhu, Ma, Yan, Jin, Li,
  and Gai}{Zhou et~al\mbox{.}}{2018}]%
        {zhou2018deep}
\bibfield{author}{\bibinfo{person}{Guorui Zhou}, \bibinfo{person}{Xiaoqiang
  Zhu}, \bibinfo{person}{Chenru Song}, \bibinfo{person}{Ying Fan},
  \bibinfo{person}{Han Zhu}, \bibinfo{person}{Xiao Ma},
  \bibinfo{person}{Yanghui Yan}, \bibinfo{person}{Junqi Jin},
  \bibinfo{person}{Han Li}, {and} \bibinfo{person}{Kun Gai}.}
  \bibinfo{year}{2018}\natexlab{}.
\newblock \showarticletitle{Deep interest network for click-through rate
  prediction}. In \bibinfo{booktitle}{\emph{Proceedings of the 24th ACM SIGKDD
  International Conference on Knowledge Discovery \& Data Mining}}. ACM,
  \bibinfo{pages}{1059--1068}.
\newblock


\end{thebibliography}

\end{document}